\documentclass[aps,prl,reprint,superscriptaddress,twocolumn,longbibliography]{revtex4-2}
\usepackage{amsfonts,amssymb,amscd,amsthm,amsmath}
\usepackage{graphicx}
\usepackage{mathrsfs}
\usepackage[colorlinks, citecolor=red]{hyperref}
\usepackage{etoolbox}
\usepackage{upgreek}

\begin{document}

\title{Observation of Non-Gaussian Magnon Dynamics in a Two-Dimensional Long-Range XY Model}

\author{S.-A. Guo}
\affiliation{Center for Quantum Information, Institute for Interdisciplinary Information Sciences, Tsinghua University, Beijing 100084, PR China}

\author{J.-Y. Tan}
\affiliation{Center for Quantum Information, Institute for Interdisciplinary Information Sciences, Tsinghua University, Beijing 100084, PR China}

\author{J. Ye}
\affiliation{Center for Quantum Information, Institute for Interdisciplinary Information Sciences, Tsinghua University, Beijing 100084, PR China}

\author{Y. Jiang}
\affiliation{Center for Quantum Information, Institute for Interdisciplinary Information Sciences, Tsinghua University, Beijing 100084, PR China}

\author{L. Zhang}
\affiliation{Center for Quantum Information, Institute for Interdisciplinary Information Sciences, Tsinghua University, Beijing 100084, PR China}

\author{Y.-X. Chen}
\affiliation{Center for Quantum Information, Institute for Interdisciplinary Information Sciences, Tsinghua University, Beijing 100084, PR China}

\author{H.-J. Chen}
\affiliation{Center for Quantum Information, Institute for Interdisciplinary Information Sciences, Tsinghua University, Beijing 100084, PR China}

\author{H.-Y. Hu}
\affiliation{Center for Quantum Information, Institute for Interdisciplinary Information Sciences, Tsinghua University, Beijing 100084, PR China}

\author{W.-X. Guo}
\affiliation{HYQ Co., Ltd., Beijing 100176, PR China}

\author{B.-X. Qi}
\affiliation{Center for Quantum Information, Institute for Interdisciplinary Information Sciences, Tsinghua University, Beijing 100084, PR China}

\author{L. He}
\affiliation{Center for Quantum Information, Institute for Interdisciplinary Information Sciences, Tsinghua University, Beijing 100084, PR China}
\affiliation{Hefei National Laboratory, Hefei 230088, PR China}

\author{Z.-C. Zhou}
\affiliation{Center for Quantum Information, Institute for Interdisciplinary Information Sciences, Tsinghua University, Beijing 100084, PR China}
\affiliation{Hefei National Laboratory, Hefei 230088, PR China}

\author{Y.-K. Wu}
\email{wyukai@mail.tsinghua.edu.cn}
\affiliation{Center for Quantum Information, Institute for Interdisciplinary Information Sciences, Tsinghua University, Beijing 100084, PR China}
\affiliation{Hefei National Laboratory, Hefei 230088, PR China}

\author{L.-M. Duan}
\email{lmduan@tsinghua.edu.cn}
\affiliation{Center for Quantum Information, Institute for Interdisciplinary Information Sciences, Tsinghua University, Beijing 100084, PR China}
\affiliation{Hefei National Laboratory, Hefei 230088, PR China}
\affiliation{New Cornerstone Science Laboratory, Institute for Interdisciplinary Information Sciences, Tsinghua University, Beijing 100084, PR China}

\begin{abstract}
Non-Gaussian evolution of high-order spin correlations characterizes important properties of quantum many-body systems. In practice, decoherence, statistical fluctuation and miscalibration of experimental parameters all hinder the witness of non-Gaussian dynamics. Here we demonstrate the crossover between Gaussian and non-Gaussian dynamics on a two-dimensional $XY$ model with long-range and spatially structured interaction using a trapped ion quantum simulator. We prepare different initial densities of magnon excitations and verify the dynamics of single-spin observables for the engineered Hamiltonian. Then we compare the high-order spin correlations with the mean-field solution and the Holstein-Primakoff approximation, and demonstrate the non-Gaussian behavior in a way independent of the calibration errors. Our work provides a verifiable path from classically simulatable dynamics to regimes where quantum advantage may emerge.
\end{abstract}

\maketitle

Simulating many-body quantum dynamics is among the most important applications of universal quantum computers and special-purpose quantum simulators \cite{RevModPhys.86.153,RevModPhys.94.015004,Daley2022}. Due to the exponential cost of quantum state tomography, it is common to characterize the properties of the simulated quantum system like phase transition \cite{sachdev2000quantum}, thermalization \cite{DAlessio03052016,Swingle2018} and multipartite entanglement \cite{GUHNE20091}, by correlations of observables in space and time \cite{doi:10.1073/pnas.37.7.452,doi:10.1073/pnas.37.7.455,Schweigler2017}. For bosonic or fermionic systems, many physically relevant states are close to the ground states or thermal states of quadratic Hamiltonians, including their time evolution under those Hamiltonians \cite{Schuch2006,WANG20071,doi:10.1142/S1230161214400010,10.21468/SciPostPhysLectNotes.54}. High-order correlators of such Gaussian states are completely determined by the first-order and second-order ones through the Wick's theorem \cite{PhysRev.80.268}, allowing efficient classical simulation of the quantum dynamics \cite{PhysRevLett.88.097904,PhysRevLett.109.230503,10.21468/SciPostPhysLectNotes.54}. Therefore, to achieve universality \cite{PhysRevLett.97.110501,RevModPhys.77.513} and quantum advantage \cite{PhysRevLett.119.170501,doi:10.1126/science.abe8770,Madsen2022,Liu2026} on a continuous-variable quantum computer, it is necessary to have non-Gaussian resource states, gates or measurements.

Different from bosons or fermions, dynamics of spin systems is intrinsically ``non-Gaussian'' even under a simple Hamiltonian with at most two-body interactions, because such terms generally do not form a closed Lie algebra, preventing efficient classical simulation \cite{PhysRevLett.97.190501}. Indeed, quantum advantage has been demonstrated in spin systems for random circuit sampling tasks \cite{arute2019quantum,PhysRevLett.127.180501,PhysRevX.15.021052} and for the measurement of multi-spin correlations \cite{Kim2023,doi:10.1126/science.ado6285} and the out-of-time-ordered correlators \cite{Abanin2025}. Nevertheless, in practice the dynamics of many spin models can be approximated efficiently by classical methods. For one-dimensional systems near the ground states, the density matrix renormalization group or the matrix product state methods are highly effective \cite{RevModPhys.77.259,SCHOLLWOCK201196}. For high-dimensional systems and long-range interactions the dynamics can often be described by the mean-field theory or the discrete truncated Wigner approximation (DTWA) with a truncation at the first or the second order of spin correlations in the equation of motion \cite{PhysRevX.5.011022,Schachenmayer_2015,PhysRevB.93.174302}. Another commonly used approximation is to map the spin interaction into a quadratic bosonic Hamiltonian through the Holstein-Primakoff (HP) transformation under a weak excitation \cite{PhysRev.58.1098,RevModPhys.63.1}. In particular, the latter two methods can be viewed as generalizations of the Gaussian theory to the spin systems. This makes it not only fundamentally interesting to witness the transition from Gaussian to non-Gaussian spin dynamics, but also as a route to demonstrating quantum advantage in practical quantum simulation tasks. However, to experimentally observe such a crossover can be challenging. On the one hand, limited by the finite accuracy due to the experimental imperfection and miscalibration, and the statistical fluctuation, the approximation methods can remain sufficient even with considerable violation of their validity conditions. On the other hand, experimentally it is common to start from product states with no spin correlations, while the late-time dynamics often leads to uncorrelated final states due to either the inevitable decoherence \cite{Sieberer_2016,PhysRevA.62.062311,PhysRevLett.129.260405} or the quantum thermalization \cite{DAlessio03052016,RevModPhys.83.863}. Hence a suitable intermediate timescale is needed for the appearance of non-Gaussian dynamics, apart from a well-controlled quantum system with a low decoherence rate. (Admittedly, for carefully engineered spin systems and environments, it is possible that the steady state can maintain nontrivial entanglement and correlations \cite{PhysRevA.78.042307,Verstraete2009}.)

\begin{figure}[!tbp]
\centering
\includegraphics[width=\linewidth]{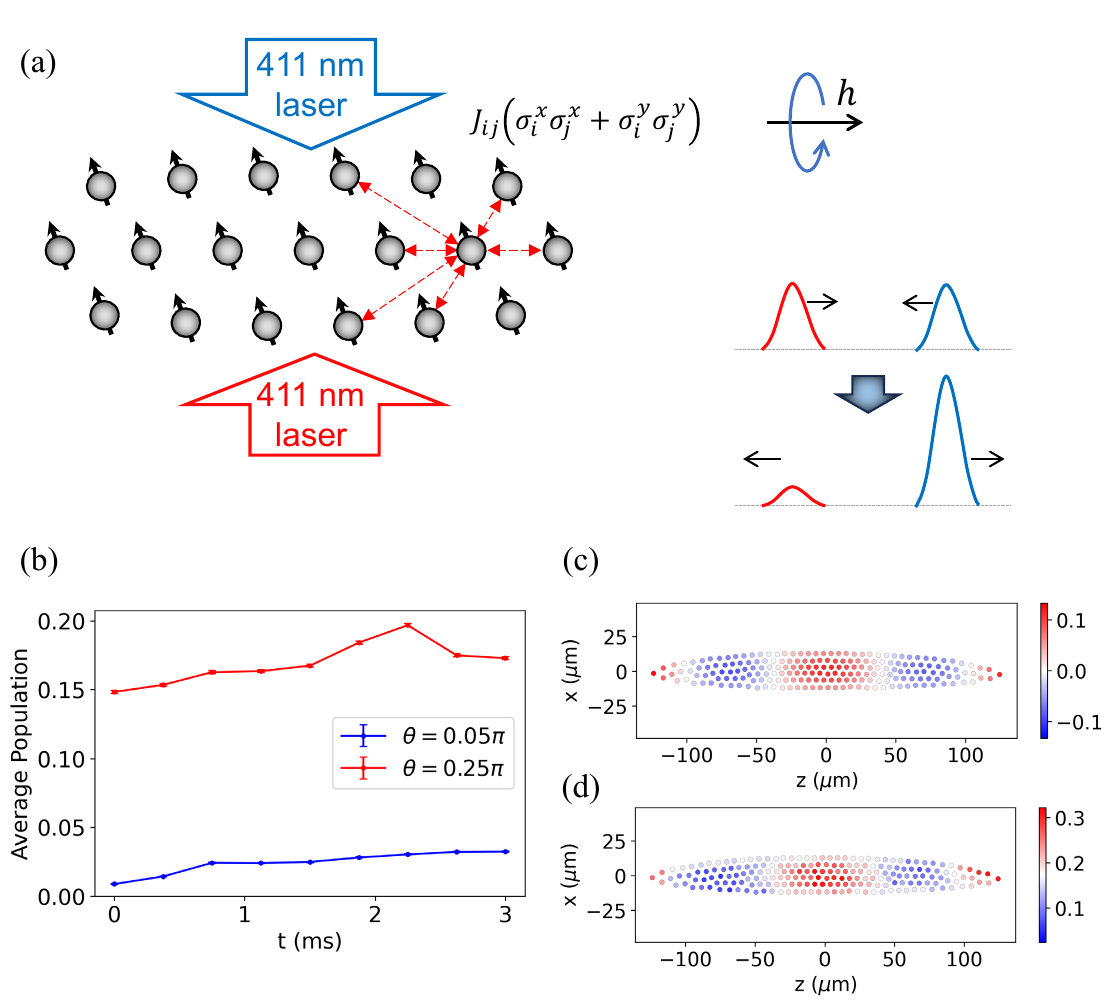}
\caption {(a) Experimental scheme. We apply counter-propagating $411\,$nm laser beams to realize a long-range and spatially structured interaction between ions in a 2D crystal, and a resonant microwave to generate a strong transverse field $h$. Under the rotating-wave approximation and a redefinition of the $x$ and $z$ directions, we get a long-range $XY$ model Hamiltonian. We initialize the spins at an angle of $\theta$ from the $z$ axis. Under the Gaussian approximation, the spin-wave excitations or magnons are non-interacting, while non-Gaussian interaction scatters the magnons into different directions with modified amplitudes in the dynamics, as sketched in the lower right panel. (b) Dynamics of the average magnon population under a weak (blue, $\theta=0.05\pi$) and a strong (red, $\theta=0.25\pi$) initial excitation. (c) Spatial distribution of the fifth highest phonon mode to which the spins are coupled \cite{HamiltonianLearning2025}.
(d) Local magnon excitation $\langle n_i \rangle = (1 - \langle Z_i \rangle)/2$ under the strong initial excitation $\theta=0.25\pi$ at the evolution time $t=3\,$ms.}
\label{fig1}
\end{figure}

Here we examine the non-Gaussian spin dynamics in a two-dimensional (2D) ion crystal with $N=200$ ions. We realize an $XY$ model Hamiltonian by engineering a long-range Ising interaction together with a strong transverse field. Then we initialize the system with different densities of magnon excitations, and measure the dynamics of multi-spin correlations using the site-resolved detection of individual spins. The single-spin observables are shown to follow the spatial pattern of the engineered Hamiltonian, consistent with the prediction of the HP transformation. For the multi-spin correlations, we compare them with the mean-field solution and the Wick's theorem expansions from the measured low-order correlations, and observe the emergence of non-Gaussian correlations beyond experimental errors. By integrating both Gaussian and non-Gaussian dynamics, our work provides a scalable platform for exploring classically intractable non-Gaussian many-body physics, while maintaining its verifiability in the Gaussian parameter regime.

Our experimental setup is illustrated in Fig.~\ref{fig1}(a), with a 2D crystal of $N=200$ ${}^{171}\mathrm{Yb}^+$ ions in a monolithic Paul trap at the cryogenic temperature of $T=6\,$K \cite{guo2024siteresolved}. We encode the qubits in the subspace spanned by the hyperfine clock states $|S_{1/2},F=0,m_F=0\rangle$ and $|S_{1/2},F=1,m_F=0\rangle$, and apply counter-propagating $411\,$nm laser beams, perpendicular to the 2D crystal, to generate a long-range Ising interaction \cite{guo2024siteresolved,HamiltonianLearning2025,PhysRevA.103.012603}. In addition, we apply a global microwave resonant to the qubit transition to provide a strong transverse field $h=2\pi\times 3.7\,$kHz. Under rotating-wave approximation, the system Hamiltonian can be described by an $XY$ model (see Supplemental Material for details)
\begin{equation}
H =\sum_{i< j} J_{ij} (\sigma_i^x \sigma_j^x + \sigma_i^y \sigma_j^y),
\label{eq:Hamiltonian}
\end{equation}
where $\sigma^{x(y,z)}$ represent the Pauli operators, and we have defined the $Z$ basis along the transverse field for simplicity. In this experiment, we couple the spins dominantly to the fifth highest phonon mode of the ion crystal with its mode structure shown in Fig.~\ref{fig1}(c) to engineer a spatial pattern of the coupling coefficients. The average hopping strength of individual spin excitations, or magnons, can be estimated by $g=\frac{1}{N}\sum_i\sqrt{\sum_{j\ne i} J_{ij}^2}=2\pi\times 0.08\,$kHz. This corresponds to a timescale of $1/g\sim 2\,$ms for which we will study the magnon dynamics and correlations. To verify the successful simulation of the desired Hamiltonian in Eq.~(\ref{eq:Hamiltonian}), we note that it conserves the total spin excitation.
We initialize the spins at an angle of $\theta$ from the $z$ axis by optical pumping into $|S_{1/2},F=0,m_F=0\rangle$ followed by a $(\pi/2-\theta)$ rotation using a microwave SK1 composite pulse \cite{PhysRevA.70.052318}. This corresponds to an average excitation probability of $\sin^2(\theta/2)$. In Fig.~\ref{fig1}(b) we plot the measured magnon population for an evolution time up to $3\,$ms from an initial state with weak (blue, $\theta=0.05\pi$) or strong (red, $\theta=0.25\pi$) excitations. Both curves are relatively flat, and the slow increase in the excitation probability is mainly due to the spin coherence time of $T_2\sim 10\,$ms. (Note that the dephasing error is mapped into a bit-flip error after the redefinition of the axes.)

Another approach to verify the simulated model and in particular the engineered coupling pattern is through the observation that, under a weak excitation $\theta\ll 1$ and to the lowest order of the HP transformation $\sigma^+ \to a$, $\sigma^-\to a^\dag$ and $\sigma^z \to 1-2a^\dag a$, the Hamiltonian in Eq.~(\ref{eq:Hamiltonian}) can be approximated by a quadratic bosonic model. Meanwhile, the initial spin state can also be approximated by the coherent states of the corresponding bosonic modes. In this way, the time evolution can be solved analytically, and we find that the final magnon excitation always imprints the spatial pattern of the collective phonon mode $b_{i}$ and the laser intensity $\Omega_i$ as $\langle\sigma_i^z(t)\rangle-\langle\sigma_i^z(0)\rangle\propto b_{i}\Omega_i$ (see Supplemental Material for details) where the initial $\langle\sigma_i^z(0)\rangle$ is a uniform distribution. As shown in Fig.~\ref{fig1}(d), this works qualitatively well even under a strong excitation $\theta=0.25\pi$ and a long evolution time $t=3\,$ms, comparing with the theoretically calculated phonon mode vector $b_{i}$ in Fig.~\ref{fig1}(c).

\begin{figure}[!tbp]
\centering
\includegraphics[width=\linewidth]{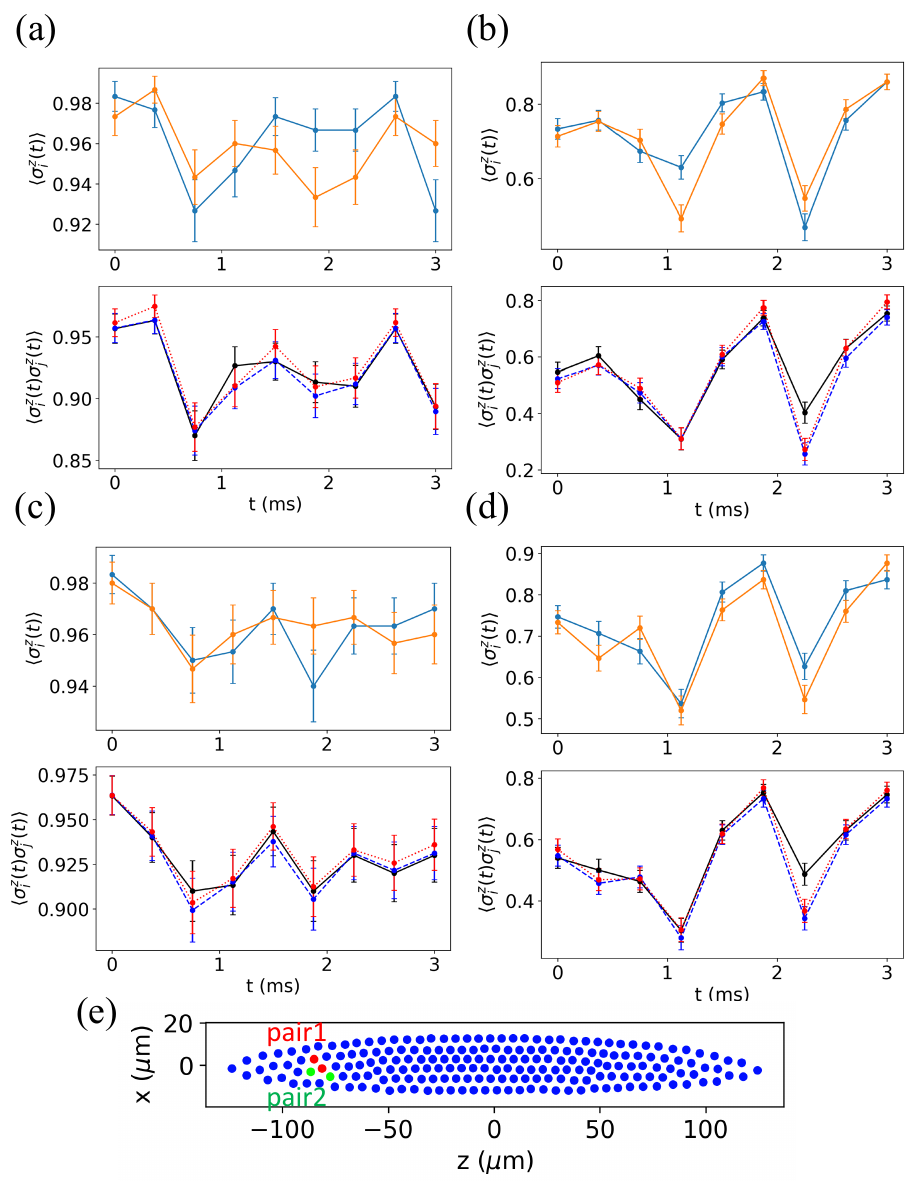}
\caption{Dynamics of $Z$-basis correlations. (a) Single-spin dynamics $\langle \sigma^z_i(t)\rangle$ (upper panel) and the two-spin correlation $\langle \sigma^z_i(t)\sigma^z_j(t)\rangle$ (lower panel) for a nearby spin pair (pair 1 in (e)) starting from a low magnon density $\theta=0.05\pi$. The measured two-spin correlation (black solid curve) is compared with the mean-field solution (blue dashed) and the Wick's theorem expansion (red dotted) calculated from low-order observables. (b) Similar plots for the spin pair 1 from a high initial magnon density $\theta=0.25\pi$. (c), (d) Similar plots for a different spin pair 2 from low and high magnon densities, respectively. Each data point is averaged over 600 samples and the error bars represent one standard error.
(e) The locations of the selected ion pairs on the 200-ion 2D crystal.
}\label{fig2}
\end{figure}

Next, we examine the violation of the Gaussian approximation in spin correlations. In Fig.~\ref{fig2} we consider the correlations in the $Z$ basis. The single-spin dynamics $\langle \sigma^z_i(t)\rangle$ for two nearby ions (pair 1 in Fig.~\ref{fig2}(e)) is shown in the upper panel of Fig.~\ref{fig2}(a) from a weakly excited initial state ($\theta=0.05\pi$), and we plot the corresponding two-spin correlation $\langle \sigma^z_i(t) \sigma^z_j(t) \rangle$ in the lower panel as the black solid curve. We compare this experimental result with two types of Gaussian approximations. The first one is simply the mean-field solution $\langle \sigma^z_i(t)\rangle \langle \sigma^z_j(t) \rangle$ (blue dashed curve) motivated by the observations that the system has long-range interaction, that it starts from a product state, and that its net correlation should decay after long-time evolution due to the decoherence. As a second approximation method, we regard the two-spin correlation in the $Z$ basis as up to four bosonic operators under the HP transformation, and expand it into low-order observables by the Wick's theorem (red dotted curve). The expansion generally requires single-qubit rotations of individual qubits in different bases, but can be approximated by global rotations and site-resolved readout as (see Supplemental Material)
\begin{equation}
\begin{aligned}
\langle \sigma^z_i \sigma^z_j \rangle \approx& \langle \sigma^z_i \rangle \langle \sigma^z_j \rangle + \frac{\langle \sigma^x_i \sigma^x_j + \sigma^y_i \sigma^y_j \rangle^2}{4\cos^2\!\left(\tan^{-1}\!\frac{\langle\sigma_i^y\rangle}{\langle\sigma_i^x\rangle} - \tan^{-1}\!\frac{\langle\sigma_j^y\rangle}{\langle\sigma_j^x\rangle}\right)} \\
& + \frac{1}{4} \left\langle \sigma^x_i \sigma^x_j - \sigma^y_i \sigma^y_j \right\rangle^2 +\frac{1}{4} \langle D_i D_j - A_i A_j \rangle^2 \\
& -\frac{1}{2} \left(\langle \sigma^x_i\rangle^2 + \langle \sigma^y_i\rangle^2 \right)\left( \langle \sigma^x_j\rangle^2 + \langle \sigma^y_j\rangle^2 \right),
\end{aligned}\label{eq:Wick}
\end{equation}
where $D\equiv (\sigma^x+\sigma^y)/\sqrt{2}$ and $A\equiv (\sigma^x-\sigma^y)/\sqrt{2}$ represent the diagonal and anti-diagonal bases, respectively. Note that both approximations are calculated from the experimentally measured low-order observables. This allows us to test the validity of the Gaussian approximations without worrying about the accuracy of calibrating the system Hamiltonian or modelling the decoherence sources.

The experimental results and the two approximations are close to each other within the statistical error bars in Fig.~\ref{fig2}(a) under a weak initial excitation, suggesting good Gaussian dynamics for the spins. In comparison, for a stronger excitation in Fig.~\ref{fig2}(b) ($\theta=0.25\pi$), we occasionally observe the violation of the Gaussian approximation beyond the experimental error bars. Similar phenomena can also be seen for another nearby ion pair (pair 2 in Fig.~\ref{fig2}(e)) as shown in Figs.~\ref{fig2}(c) and (d). Combining Figs.~\ref{fig2}(b) and (d), we also observe an oscillatory behavior between the violation and the recovery of the Gaussian approximation. For the initial product state there is no net correlation, and then the non-Gaussianness can accumulate as the magnon excitations hop and redistribute over the 2D crystal at a timescale governed by the hopping strength $g=2\pi\times 0.08\,$kHz. Under the Gaussian approximation and the assumption of a low-rank coupling matrix $J$, we can actually predict a periodic dynamics (see Supplemental Material), and hence expect a periodic collapse and revival of the Gaussianness. Finally, we note that the difference between the two approximation methods is small even for a strong excitation, suggesting a weak accumulation of second-order correlation in the $Z$ basis and hence only weak non-Gaussianness. This motivates us to examine higher-order correlations and other measurement bases.

\begin{figure}[!tbp]
\centering
\includegraphics[width=\linewidth]{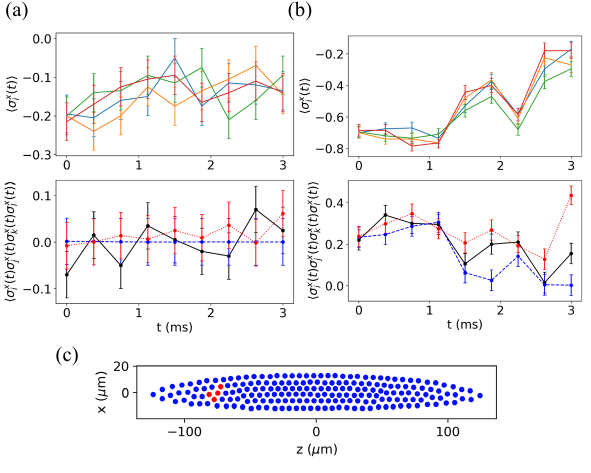}
\caption{Dynamics of $X$-basis correlations. (a) Single-spin dynamics $\langle \sigma^x_i(t)\rangle$ (upper panel) and the four-spin correlation $\langle \sigma^x_i(t)\sigma^x_j(t)\sigma^x_k(t)\sigma^x_l(t)\rangle$ (lower panel) for a nearby group of four spins as shown in (c) starting from a low magnon density $\theta=0.05\pi$. Again we compare the measured four-spin correlation (black solid) with the mean-field solution (blue dashed) and the Wick's theorem expansion (red dotted) calculated from the single-spin and two-spin observables. (b) Similar plots from a high initial magnon density $\theta=0.25\pi$. Each data point is averaged over 400 samples and the error bars represent one standard error.
(c) The locations of the selected ions on the 200-ion 2D crystal.
}\label{fig3}
\end{figure}

In Fig.~\ref{fig3} we consider the high-order spin correlations in the $X$ basis. Here we choose a group of four neighboring ions and measure their single-spin dynamics in the upper panels of Fig.~\ref{fig3}(a) for a weak excitation $\theta=0.05\pi$, and Fig.~\ref{fig3}(b) for a strong excitation $\theta=0.25\pi$. Again, we present the measured correlations in the lower panels (black solid), and compare with the mean-field approximation (blue dashed) and the Wick's theorem expansion (red dotted). Here the four-spin correlation can be viewed as four bosonic operators under the HP transformation, and can be decomposed into the measured single-spin and two-spin observables in the $X$ basis (see Supplemental Material). Similar to the $Z$ basis, we observe Gaussian dynamics within the experimental error bars under a low magnon density, and obtain alternating violation and recovery of the Gaussian approximation under a high magnon density. This time we have larger difference between the two approximations, indicating stronger accumulation of spin correlations. This can also be supported by the larger difference between the experimental data and the Wick's theorem expansion results at the late dynamics. However, the higher-order $X$-basis correlations are also subjected to stronger spin dephasing, resulting in the experimental data to be closer to the mean-field results. This again suppresses the non-Gaussianness after a long evolution time.

\begin{figure}[!tbp]
	\centering
	\includegraphics[width=\linewidth]{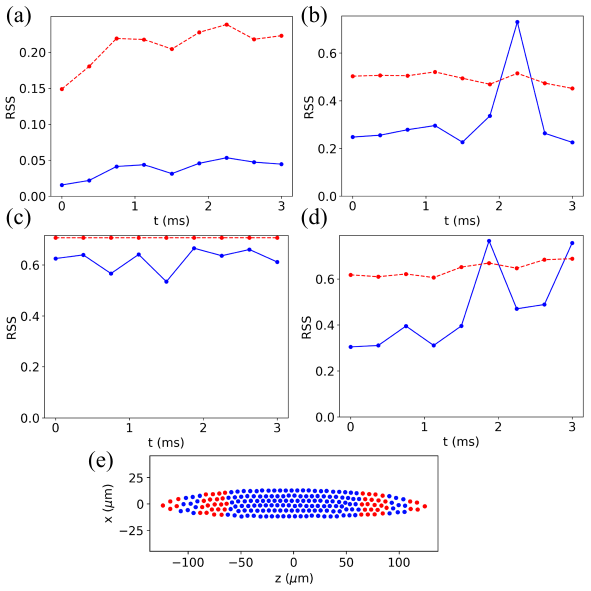}
	\caption{(a-d) Total deviation from Gaussian dynamics under a weak (the first column) or a strong (the second column) magnon density in the two-body $Z$-basis correlations (the first row) or the four-body $X$-basis correlations (the second row). The blue solid curves show the RSS of the distance between the measured correlations and the approximations, and the red dashed curves are the RSS of the error bars. (e) Selected ions (colored in red) with the highest participation in the fifth highest phonon mode, for which we compute the correlations (see the main text for details).
}\label{fig4}
\end{figure}

Finally, we examine the reliability of the demonstrated violation of the Gaussian approximation. Note that, even for the suitable initial state and suitable evolution time in Figs.~\ref{fig2} and \ref{fig3}, the non-Gaussian dynamics may not be observed for all the subgroups of spins. For example, for the nonuniform hopping rates governed by the spatial structure of the collective phonon mode as shown in Fig.~\ref{fig1}(c) (see Supplemental Material for details), the ions near the node of the phonon mode ($b_i\approx 0$, indicated by the white color) will be coupled only weakly to the others, resulting in a low correlation. On the other hand, consider a hypothetical system whose spin dynamics is completely Gaussian. We still expect to observe about $32\%$ violations beyond one standard error among the $O(N^k)$ possible $k$-body correlations of the $N$ ions. Then this leads to the question whether the experimentally observed violation is physical, or whether it just comes from the statistical fluctuation. To elucidate this matter, in Fig.~\ref{fig4} we choose the top $50$ ions with the highest theoretical participation in the phonon mode we use, and select $200$ neighboring ion pairs (4-ion subgroups) within them for which we compute the two-spin $Z$-basis (four-spin $X$-basis) correlations. For each time point, we compute the root-sum-square (RSS) of the violation (the minimal distance between the measured correlation and the two approximations) over all the selected correlations as the blue solid curve. This is to be compared with the RSS of the error bars of individual correlations as the red dashed curve. As shown in Fig.~\ref{fig4}(a-d), again we get good Gaussian approximation under a weak initial excitation, and occasional violation under a strong excitation. Note that, here by taking the RSS of both the data and the error bars, we do not change the probability that the statistical fluctuation can exceed one standard error. What improves the reliability of the result is that we reduce the number of correlations to be considered by focussing on the strongly coupled ions, such that the number of occurrence of the rare events can be largely suppressed.

To sum up, in this work we demonstrate the crossover from Gaussian to non-Gaussian quantum dynamics in a long-range $XY$ model. We compare the experimentally measured spin correlations with the mean-field approximation and the HP transformation followed by the Wick's theorem expansion. Our analysis uses solely experimentally measured dynamics and is thus independent of the miscalibration of the Hamiltonian or the decoherence error. This observed crossover of the same experimental system between Gaussian and non-Gaussian regimes provides a useful way to verify the successful quantum simulation of the model in one parameter regime, while exploring non-Gaussian dynamics beyond the capability of classical computers in the other regime.

\begin{acknowledgements}
We thank Immanuel Bloch for helpful discussions. This work was supported by Quantum Science and Technology-National Science and Technology Major Project (Grant No. 2021ZD0301601,2021ZD0301605), Beijing Science and Technology Planning Project (Grant No. Z25110100040000), the National Natural Science Foundation of China (Grant No. 12575021 and 12574541), Tsinghua University Initiative Scientific Research Program, and the Ministry of Education of China. L.M.D. acknowledges in addition support from the New Cornerstone Science Foundation through the New Cornerstone Investigator Program. Y.K.W. acknowledges in addition support from the Dushi program from Tsinghua University.

The data that support the findings of this article are openly available \cite{data}.
\end{acknowledgements}

%

\end{document}


\title{Supplemental Material for ``Observation of Non-Gaussian Magnon Dynamics in a Two-Dimensional Long-Range XY Model''}

\author{S.-A. Guo}
\affiliation{Center for Quantum Information, Institute for Interdisciplinary Information Sciences, Tsinghua University, Beijing 100084, PR China}

\author{J.-Y. Tan}
\affiliation{Center for Quantum Information, Institute for Interdisciplinary Information Sciences, Tsinghua University, Beijing 100084, PR China}

\author{J. Ye}
\affiliation{Center for Quantum Information, Institute for Interdisciplinary Information Sciences, Tsinghua University, Beijing 100084, PR China}

\author{Y. Jiang}
\affiliation{Center for Quantum Information, Institute for Interdisciplinary Information Sciences, Tsinghua University, Beijing 100084, PR China}

\author{L. Zhang}
\affiliation{Center for Quantum Information, Institute for Interdisciplinary Information Sciences, Tsinghua University, Beijing 100084, PR China}

\author{Y.-X. Chen}
\affiliation{Center for Quantum Information, Institute for Interdisciplinary Information Sciences, Tsinghua University, Beijing 100084, PR China}

\author{H.-J. Chen}
\affiliation{Center for Quantum Information, Institute for Interdisciplinary Information Sciences, Tsinghua University, Beijing 100084, PR China}

\author{H.-Y. Hu}
\affiliation{Center for Quantum Information, Institute for Interdisciplinary Information Sciences, Tsinghua University, Beijing 100084, PR China}

\author{W.-X. Guo}
\affiliation{HYQ Co., Ltd., Beijing 100176, PR China}

\author{B.-X. Qi}
\affiliation{Center for Quantum Information, Institute for Interdisciplinary Information Sciences, Tsinghua University, Beijing 100084, PR China}

\author{L. He}
\affiliation{Center for Quantum Information, Institute for Interdisciplinary Information Sciences, Tsinghua University, Beijing 100084, PR China}
\affiliation{Hefei National Laboratory, Hefei 230088, PR China}

\author{Z.-C. Zhou}
\affiliation{Center for Quantum Information, Institute for Interdisciplinary Information Sciences, Tsinghua University, Beijing 100084, PR China}
\affiliation{Hefei National Laboratory, Hefei 230088, PR China}

\author{Y.-K. Wu}
\email{wyukai@mail.tsinghua.edu.cn}
\affiliation{Center for Quantum Information, Institute for Interdisciplinary Information Sciences, Tsinghua University, Beijing 100084, PR China}
\affiliation{Hefei National Laboratory, Hefei 230088, PR China}

\author{L.-M. Duan}
\email{lmduan@tsinghua.edu.cn}
\affiliation{Center for Quantum Information, Institute for Interdisciplinary Information Sciences, Tsinghua University, Beijing 100084, PR China}
\affiliation{Hefei National Laboratory, Hefei 230088, PR China}
\affiliation{New Cornerstone Science Laboratory, Beijing 100084, PR China}

\maketitle

\section{Experimental setup}

\begin{figure}[htbp]
\centering
\includegraphics[width=0.9\linewidth]{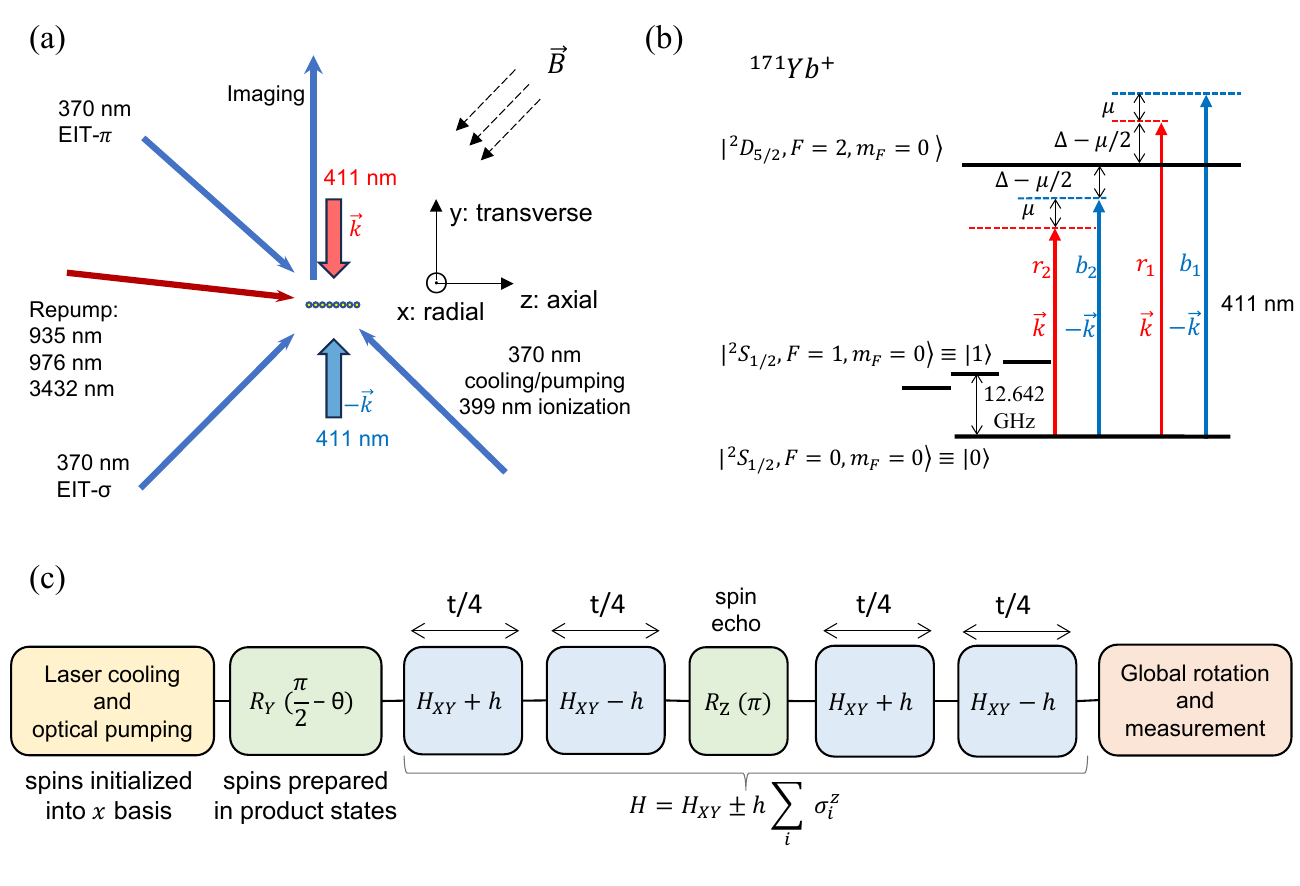}
\caption{(a) Experimental setup and laser configuration. (b) Relevant energy levels of the ${}^{171}\mathrm{Yb}^+$ ion and the frequency components and the wave vectors of the 411 nm laser beams. (c) Experimental sequence.} \label{figS:setup}
\end{figure}
Our experimental setup is detailed in Fig.~\ref{figS:setup}(a). We use a monolithic Paul trap at the temperature of $T=6\,$K to confine a 2D ion crystal \cite{guo2024siteresolved,HamiltonianLearning2025} with $N=200$ ions.
The 2D crystal is located on the $xz$ plane where $z$ is the axial direction and $x$ is one of the radial directions, with a magnetic field of $4.5\,$Gauss applied at $45^\circ$ to the crystal plane. The trap frequencies are set to $(\omega_x,\,\omega_y,\,\omega_z) = 2\pi\times (0.53,\,1.96,\,0.11)\,$MHz. A global $370\,$nm laser beam is shined on the ions along the micromotion-free $yz$ plane at an angle to both axes, serving for Doppler cooling, optical pumping, and state detection. Two additional blue-detuned $370\,$nm laser beams are configured with their wave vector difference perpendicular to the 2D ion crystal, providing EIT cooling for the transverse phonon modes. We further use a $411\,$nm laser for sideband cooling and electron shelving \cite{Roman2020,edmunds2020scalable}. A $935\,$nm laser beam is used to repump the population in the $D_{3/2}$ levels, and is combined with $976\,$nm and $3432\,$nm lasers which repump the population in the $D_{5/2}$ and $F_{7/2}$ levels.

Fig.~\ref{figS:setup}(b) shows the relevant energy levels to generate the long-range spin-spin coupling (see Sec.~\ref{sec:Gaussian}). The qubits are encoded in the ${}^{2}S_{1/2}$ hyperfine clock states of the ${}^{171}\mathrm{Yb}^+$ ions. A resonant microwave at the frequency of $12.642\,$GHz can provide global rotations. The spin-dependent force is achieved by counter-propagating off-resonant $411\,$nm laser beams \cite{guo2024siteresolved, HamiltonianLearning2025}. To to maximize the coupling strength between the $|S_{1/2}, F=0, m_F=0\rangle$ and the $|D_{5/2}, F=2, m_F=0\rangle$ states, we set the polarizations of these $411\,$nm laser beams within the plane spanned by the magnetic field and the laser beams. Each laser beam contains two frequency components on the two sides of the $S$-$D$ transition to compensate their time-independent AC Stark shift on the $|S_{1/2},F=0,m_F=0\rangle$ level.

The experimental sequence is shown in Fig.~\ref{figS:setup}(c). After laser cooling and optical pumping, we prepare all the spins at an angle of $\theta$ from the $x$ axis (which is later redefiend as the $z$ axis as described in Sec.~\ref{sec:Gaussian}) by a microwave SK1 composite pulse.
Then we turn on the desired Hamiltonian for a controllable time $t$. A spin echo is inserted in the middle to cancel the residual AC Stark shift, and we switch the direction of the transverse field during the sequence to compensate its nonuniformity (see Sec.~\ref{sec:Gaussian}). Finally, we apply another microwave SK1 composite pulse to rotate the spins into any desired basis for individual state readout by electron shelving.

\section{System Hamiltonian and Gaussian Approximation}
\label{sec:Gaussian}
Following Ref.~\cite{guo2024siteresolved}, we apply counter-propagating $411\,$nm laser beams perpendicular to the 2D ion crystal to generate a spin-dependent force. The relevant energy levels and the frequency components of the $411\,$nm laser are sketched in Fig.~\ref{figS:setup}(b) for completeness. If we set the laser detuning $\mu$ close to a collective phonon mode with a frequency $\omega_k$, a Lamb-Dicke parameter $\eta_k$ and a normalized mode vector $b_{ik}$, while still maintaining virtual excitation of the phonon mode, we obtain the Ising-type interaction
\begin{equation}
H_{\mathrm{Ising}} = \sum_{ij} \frac{\Omega_i \Omega_j}{4} \frac{\eta_k^2 b_{ik} b_{jk}}{\mu-\omega_k} (|0\rangle_i\langle 0|) (|0\rangle_j\langle 0|)\equiv \sum_{ij} J_{ij} (I+\sigma^z_i) (I+\sigma^z_j), \label{eq:Ising}
\end{equation}
where $\Omega_i$ is the $411\,$nm-laser-induced light shift on the $|0\rangle\equiv |S_{1/2},F=0,m_F=0\rangle$ level of ion $i$ ($i=1,\,2,\,\cdots,\,N$), and the coupling coefficients $J_{ij}$ are given by
\begin{equation}
J_{ij} = \frac{\eta_k^2 b_{ik} b_{jk} \Omega_i \Omega_j}{16(\mu-\omega_k)}. \label{eq:J}
\end{equation}

Then we apply a strong transverse field $h$ by a global microwave resonant to the transition between $|0\rangle$ and $|1\rangle\equiv |S_{1/2},F=1,m_F=0\rangle$ and obtain the total Hamiltonian
\begin{equation}
H = \sum_{ij} J_{ij} (I+\sigma^x_i) (I+\sigma^x_j) + h \sum_i \sigma^z_i, \label{eq:total}
\end{equation}
where we first define the $X$ axis by the phase of the microwave, and then exchange the $X$ and $Z$ bases to match the convention of the $XY$ model Hamiltonian which we want to derive. Finally we move into an interaction picture of the transverse field $H_0=h \sum_i \sigma^z_i$, and perform rotating-wave approximation to drop high-frequency terms. This leads us to the $XY$ model
\begin{equation}
H_{XY} = \sum_{ij} J_{ij} (\sigma^+_i \sigma^-_j + \sigma^+_j \sigma^-_i)=\frac{1}{2}\sum_{ij} J_{ij} (\sigma^x_i \sigma^x_j + \sigma^y_i \sigma^y_j). \label{eq:XY}
\end{equation}
Note that this expression differs from that in the main text by a coefficient of $1/2$ because here we are summing over all ion pairs $i\ne j$ while in the main text we consider $i<j$. Also note that this Hamiltonian is obtained in a frame rotating at the frequency of $h$ with respect to the lab frame, and in principle we need to compensate this rotation before we measure any observables in the $X$ or $Y$ bases. However, as shown in Fig.~\ref{figS:setup}(c) for the experimental sequence, we alternatingly apply the transverse field of $+h$ and $-h$ to compensate any nonuniformity of the microwave across the large ion crystal. Therefore, after the whole experimental sequence, the rotating frame and the lab frame coincide again, and there is no need to compensate additional rotations.

Under the assumption of a weak average excitation, we take the Holstein-Primakoff transformation to the lowest order by mapping $\sigma^+ \to a$, $\sigma^-\to a^\dag$ and $\sigma^z \to 1-2a^\dag a$, and get a quadratic bosonic Hamiltonian
\begin{equation}
H_b = \sum_{ij} J_{ij} (a^\dag_i a_j + a^\dag_j a_i).
\end{equation}
We can further diagonalize the symmetric coupling matrix $J$ by a unitary $U$ such that $J=U\Lambda U^\dag$ where $\Lambda=\mathrm{diag}\{\lambda_1,\,\cdots,\,\lambda_N\}$ is a diagonal matrix. Then we can express the Hamiltonian as
\begin{equation}
H_b = 2\sum_i \lambda_i b_i^\dag b_i,
\end{equation}
where the transformed modes are related to the original modes by $b_i\equiv \sum_j u_{ji}^* a_j$ and $a_j=\sum_i u_{ji} b_i$.

As described in the main text, we initialize each spin at an angle of $\theta$ from the $Z$ direction. This corresponds to a coherent state $|\psi_0\rangle=\otimes_{i=1}^N |\alpha_i=\sin(\theta/2)\rangle$ under the Gaussian approximation in terms of the original modes. In the following, we denote such a coherent state as $|\boldsymbol{\alpha}\rangle\equiv \otimes_{i=1}^N |\alpha_i\rangle$ for simplicity. In terms of the transformed modes, the initial state can also be expressed as $|\psi_0\rangle=|\boldsymbol{\beta}\rangle$ where $\beta_i=\sum_j u_{ji}^*\sin(\theta/2)$.
With these notations, the final state after an evolution time $t$ can be expressed as $|\boldsymbol{\beta}(t)\rangle$ where $\beta_i(t)=e^{-2i\lambda_i t}\beta_i(0)$. Transforming back to the original bosonic modes, we thus have the final state as $|\boldsymbol{\alpha}(t)\rangle$ where $\boldsymbol{\alpha}(t)=U e^{-2i\Lambda t} U^\dag \boldsymbol{\alpha}(0)$.

While the above derivations work for general coupling matrices $J$, here we focus on the special case that $J$ is generated by coupling to a specific collective phonon mode as described in Eq.~(\ref{eq:J}). In this situation, we have the coupling matrix $J$ close to a rank-1 matrix $J\approx \lambda \boldsymbol{u} \boldsymbol{u}^T$ (we will discuss the small correction later) where $\boldsymbol{u}$ is a normalized column vector satisfying $u_i= c b_{ik}\Omega_i$, and $\lambda=\eta_k^2/[16c^2(\mu-\omega_k)]$. Then we have $\boldsymbol{\alpha}(t)=[I+(e^{-2i\lambda t}-1) \boldsymbol{u} \boldsymbol{u}^T] \boldsymbol{\alpha}(0)$ which has a period of $\pi/\lambda$. A direct observation is that the difference between the initial and the final states is always proportional to the vector $\boldsymbol{u}$ and hence imprints the collective phonon mode $b_{ik}$ together with the distribution of the laser intensity $\Omega_i$. When transforming back to the spin observables, we have $\delta\langle \sigma_i^x\rangle, \delta\langle\sigma_i^y \rangle \propto u_i$ for the change in the single-spin $\sigma_i^x$ and $\sigma_i^y$ observables. As for the change in the $\sigma_i^z$ observables, we also have $\delta\langle\sigma_i^z\rangle\propto u_i$ for a short evolution time or for a small overlap between $\boldsymbol{u}$, which is close to the mode vector $b_{ik}$ of the fifth highest phonon mode, and $\boldsymbol{\alpha}(0)$, which is uniform and hence is close to the mode vector of the center-of-mass mode. This explains the observed spatial pattern in Fig.~1(e) of the main text.

Finally, we note that the coupling matrix in Eq.~(\ref{eq:J}) is not exactly a rank-1 matrix, because that expression only holds for $i\ne j$ while we need to set $J_{ii}=0$ ($i=1,\,2,\,\cdots,\,N$). Specifically, we have $J=\lambda \boldsymbol{u} \boldsymbol{u}^T-\lambda \cdot \mathrm{diag}\{u_i^2\}$. Nevertheless, under the long-range coupling between the ions, the typical hopping strength for the ion $i$ can be given by $\sqrt{\sum_{j\ne i} J_{ij}^2}\sim \sqrt{N}J_0$ where we define $J_0=\sqrt{\langle J_{ij}^2\rangle}$, while the diagonal term is only of the order $J_{ii}\sim J_0$. Therefore, for the timescale in this experiment to observe the hopping of the magnon excitations, the additional phase caused by the diagonal term will not be significant.

\section{Decomposition of High-order Correlations by Observables based on Global Rotations}
\label{sec:Wick}
In this section we describe how we approximate the higher-order spin correlations by the Holstein-Primakoff transformation $\sigma^+ \leftrightarrow a$, $\sigma^-\leftrightarrow a^\dag$ and $\sigma^z \leftrightarrow 1-2a^\dag a$.

For the $Z$-basis correlation, we have $\sigma^z_i \sigma^z_j \to (1-2a_i^\dag a_i)(1-2a_j^\dag a_j)=1-2a_i^\dag a_i-2a_j^\dag a_j+4a_i^\dag a_i a_j^\dag a_j$. Then we use the Wick's theorem to expand the last term into the first-order and the second-order terms, because as we describe in Sec.~\ref{sec:Gaussian}, the evolved state is a Gaussian state under a weak excitation.
\begin{equation}
\langle a_i^\dag a_i a_j^\dag a_j\rangle = \langle a_i^\dag a_i\rangle \langle a_j^\dag a_j\rangle + \langle a_i^\dag a_j\rangle \langle a_j^\dag a_i\rangle + \langle a_i^\dag a_j^\dag\rangle \langle a_i a_j\rangle - 2 \langle a_i^\dag \rangle\langle a_i\rangle \langle a_j^\dag \rangle\langle a_j\rangle.
\end{equation}
Therefore, we have
\begin{equation}
\begin{aligned}
\left\langle \sigma^z_i \sigma^z_j \right\rangle \approx& \left\langle \sigma^z_i \right\rangle + \left\langle \sigma^z_j \right\rangle - 1 + \left(1 - \left\langle \sigma^z_i \right\rangle\right) \left(1 - \left\langle \sigma^z_j \right\rangle\right) + 4 \left\langle\sigma_i^+\sigma_j^-\right\rangle \left\langle\sigma_j^+\sigma_i^-\right\rangle + 4 \left\langle\sigma_i^+\sigma_j^+\right\rangle \left\langle\sigma_i^-\sigma_j^-\right\rangle \\
&- 8 \left\langle\sigma_i^+\right\rangle \left\langle\sigma_i^-\right\rangle \left\langle\sigma_j^+\right\rangle \left\langle\sigma_j^-\right\rangle \\
=& \left\langle \sigma^z_i \right\rangle \left\langle \sigma^z_j \right\rangle + \frac{1}{4}\left\langle\left(\sigma^x_i+i\sigma^y_i\right)\left(\sigma^x_j-i\sigma^y_j\right)\right\rangle \left\langle\left(\sigma^x_i-i\sigma^y_i\right)\left(\sigma^x_j+i\sigma^y_j\right)\right\rangle \\
& + \frac{1}{4}\left\langle\left(\sigma^x_i+i\sigma^y_i\right)\left(\sigma^x_j+i\sigma^y_j\right)\right\rangle \left\langle\left(\sigma^x_i-i\sigma^y_i\right)\left(\sigma^x_j-i\sigma^y_j\right)\right\rangle - \frac{1}{2}\left\langle\sigma^x_i+i\sigma^y_i\right\rangle \left\langle\sigma^x_i-i\sigma^y_i\right\rangle \left\langle\sigma^x_j+i\sigma^y_j\right\rangle \left\langle\sigma^x_j-i\sigma^y_j\right\rangle \\
\approx & \left\langle \sigma^z_i \right\rangle \left\langle \sigma^z_j \right\rangle + \frac{1}{4} \left\langle \sigma^x_i \sigma^x_j + \sigma^y_i \sigma^y_j \right\rangle^2 + \frac{1}{4} \left\langle \sigma^x_i \sigma^x_j - \sigma^y_i \sigma^y_j \right\rangle^2 +\frac{1}{4} \left\langle D_i D_j - A_i A_j \right\rangle^2 \\
& -\frac{1}{2} \left(\left\langle \sigma^x_i\right\rangle^2 + \left\langle \sigma^y_i\right\rangle^2 \right)\left( \left\langle \sigma^x_j\right\rangle^2 + \left\langle \sigma^y_j\right\rangle^2 \right)\\
=& \left\langle \sigma^z_i \right\rangle \left\langle \sigma^z_j \right\rangle + \frac{1}{2} \left(\left\langle \sigma^x_i \sigma^x_j \right\rangle^2 + \left\langle \sigma^y_i \sigma^y_j \right\rangle^2\right) +\frac{1}{4} \left(\left\langle D_i D_j \right\rangle - \left\langle A_i A_j \right\rangle\right)^2 -\frac{1}{2} \left(\left\langle \sigma^x_i\right\rangle^2 + \left\langle \sigma^y_i\right\rangle^2 \right)\left( \left\langle \sigma^x_j\right\rangle^2 + \left\langle \sigma^y_j\right\rangle^2 \right),
\end{aligned}\label{eq:Wick}
\end{equation}
where $D\equiv (\sigma^x+\sigma^y)/\sqrt{2}$ and $A\equiv (\sigma^x-\sigma^y)/\sqrt{2}$ are Pauli operators in the diagonal and off-diagonal bases. In the second approximation we discard the term $\frac{1}{4}\langle \sigma^x_i \sigma^y_j - \sigma^y_i \sigma^x_j \rangle^2$ which requires measuring different spins in different bases. This corresponds to discarding the imaginary part of $\langle a_i^\dag a_j\rangle$ and $\langle a_j^\dag a_i\rangle$. This approximation can be justified by the derivation in Sec.~\ref{sec:Gaussian} that, under the Gaussian approximation, the evolved state is given by $\boldsymbol{\alpha}(t)=[I+(e^{-2i\lambda t}-1) \boldsymbol{u} \boldsymbol{u}^T] \boldsymbol{\alpha}(0)$. In the main text we consider the correlation between nearby ions, and we have a smooth distribution of the phonon mode vector in space as shown in Fig.~1(c). Therefore we expect the elements $u_i$ and $u_j$ to be close to each other, hence the final coherent state $\alpha_i(t)$ and $\alpha_j(t)$. Then we have $\langle a_j^\dag a_i\rangle\approx \alpha_j^*(t)\alpha_i(t)$ to be close to a real number.

\begin{figure}[!tbp]
\centering
\includegraphics[width=\linewidth]{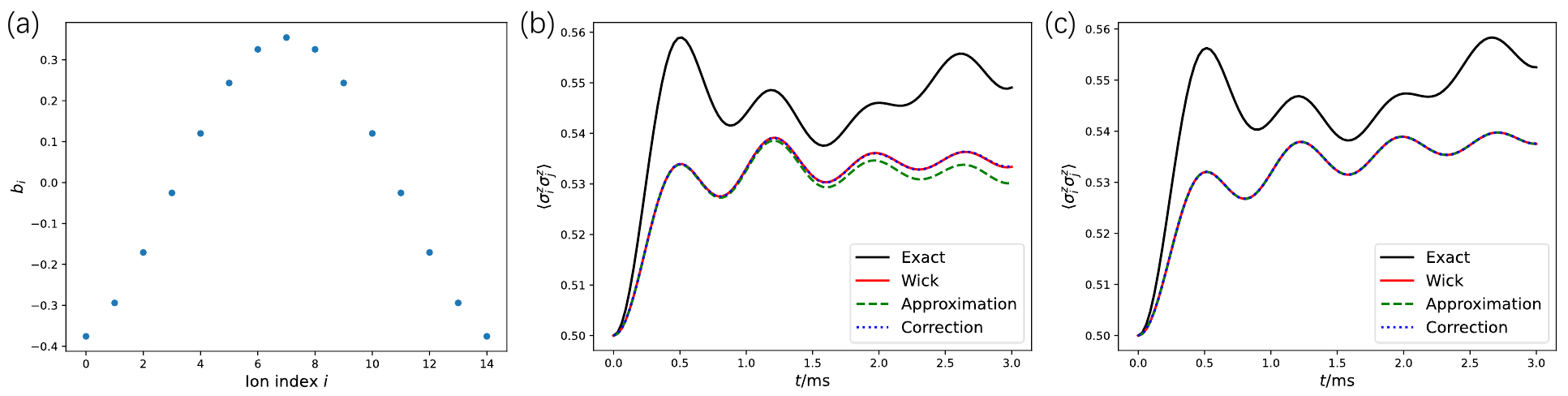}
\caption{We consider a $15$-ion chain with equal spacing $d=5\,\upmu$m, and solve its collective phonon modes under a radial trap frequency of $\omega_x=2\pi\times 3\,$MHz. The third highest mode is visualized in (a), which is symmetric and has an anti-node at the ion~$7$. We further compute the coupling matrix $J_{ij}$ following Eq.~(\ref{eq:J}), and rescale the coefficients to get an average hopping rate $g=2\pi\times 0.08\,$kHz, similar to the case in the main text. (b) Dynamics of the two-spin correlation $\left\langle \sigma^z_6 \sigma^z_7 \right\rangle$ for two neighboring ions with similar mode participation. The exact solution is shown as the black solid curve, and the Wick's theorem expansion (the first line of Eq.~(\ref{eq:Wick})) as the red solid curve. The approximate results in Eq.~(\ref{eq:Wick}) and Eq.~(\ref{eq:Wick2}) are plotted as the green dashed curve and the blue dotted curve, respectively. (c) Similar plot for the non-neighboring but symmetric ion pair $6$ and $8$. In this case the two approximations seem to be perfect.
}\label{figs:numerical}
\end{figure}

As a further correction, we can assume that $\langle a_j^\dag a_i\rangle$ has the same complex angle as $\langle a_j^\dag\rangle\langle a_i\rangle$. This leads us to
\begin{equation}
\begin{aligned}
\left\langle \sigma^z_i \sigma^z_j \right\rangle \approx& \left\langle \sigma^z_i \right\rangle \left\langle \sigma^z_j \right\rangle + \frac{\left\langle \sigma^x_i \sigma^x_j + \sigma^y_i \sigma^y_j \right\rangle^2}{4\cos^2\left(\tan^{-1}\frac{\left\langle\sigma_i^y\right\rangle}{\left\langle\sigma_i^x\right\rangle} - \tan^{-1}\frac{\left\langle\sigma_j^y\right\rangle}{\left\langle\sigma_j^x\right\rangle}\right)} + \frac{1}{4} \left\langle \sigma^x_i \sigma^x_j - \sigma^y_i \sigma^y_j \right\rangle^2 +\frac{1}{4} \left\langle D_i D_j - A_i A_j \right\rangle^2 \\
& -\frac{1}{2} \left(\left\langle \sigma^x_i\right\rangle^2 + \left\langle \sigma^y_i\right\rangle^2 \right)\left( \left\langle \sigma^x_j\right\rangle^2 + \left\langle \sigma^y_j\right\rangle^2 \right).
\end{aligned}\label{eq:Wick2}
\end{equation}
We present a numerical simulation in Fig.~\ref{figs:numerical} for a small-scale system under realistic parameters, and find that Eq.~(\ref{eq:Wick2}) gives almost identical result as the exact expansion of the Wick's theorem (the first line in Eq.~(\ref{eq:Wick})).

As for the $X$-basis correlation, we directly have
\begin{equation}
\begin{aligned}
\left\langle\sigma_i^x\sigma_j^x\sigma_k^x\sigma_l^x\right\rangle \approx & \, 4 \left\langle\sigma_i^x\sigma_j^x\right\rangle\left\langle\sigma_k^x\sigma_l^x\right\rangle + 4 \left\langle\sigma_i^x\sigma_k^x\right\rangle\left\langle\sigma_j^x\sigma_l^x\right\rangle + 4 \left\langle\sigma_i^x\sigma_l^x\right\rangle\left\langle\sigma_j^x\sigma_k^x\right\rangle  \\
&- 3 \left\langle\sigma_i^x\sigma_j^x\right\rangle\left\langle\sigma_k^x\right\rangle\left\langle\sigma_l^x\right\rangle
- 3 \left\langle\sigma_i^x\sigma_k^x\right\rangle\left\langle\sigma_j^x\right\rangle\left\langle\sigma_l^x\right\rangle
- 3 \left\langle\sigma_i^x\sigma_l^x\right\rangle\left\langle\sigma_j^x\right\rangle\left\langle\sigma_k^x\right\rangle \\
&- 3 \left\langle\sigma_j^x\sigma_k^x\right\rangle\left\langle\sigma_i^x\right\rangle\left\langle\sigma_l^x\right\rangle
- 3 \left\langle\sigma_j^x\sigma_l^x\right\rangle\left\langle\sigma_i^x\right\rangle\left\langle\sigma_k^x\right\rangle
- 3 \left\langle\sigma_k^x\sigma_l^x\right\rangle\left\langle\sigma_i^x\right\rangle\left\langle\sigma_j^x\right\rangle \\
&+ 7 \left\langle\sigma_i^x\right\rangle\left\langle\sigma_j^x\right\rangle\left\langle\sigma_k^x\right\rangle\left\langle\sigma_l^x\right\rangle,
\end{aligned}
\end{equation}
where only requires measurements in a single basis.

\bibliography{reference}